\documentstyle[12pt]{article}
\catcode`\@=11
\begin{document} \openup6pt

\title{PROBABILITY FOR PRIMORDIAL BLACK HOLES IN HIGHER DIMENSIONAL
        UNIVERSE}

\author{B. C. Paul\thanks{ e-mail : bcpaul@nbu.ernet.in } \\
   Department of Physics, North Bengal University, \\
   Siliguri, Dist. Darjeeling, Pin : 734 430, INDIA }

\date{}

\maketitle

\begin{abstract}

We investigate higher dimensional cosmological
models in the semiclassical approximation with Hartle-Hawking Boundary 
conditions, assuming a gravitational action 
which is described by  the scalar curvature with a cosmological constant.
In the framework the probability for quantum creation of an inflationary
universe with a pair of black holes in a multidimensional
universe is evaluated. 
The probability for creation of a universe with 
a spatial section with $S^{1}XS^{D -2}$ topology is then compared with that of
a higher dimensional de Sitter universe with $S^{D -1}$ spatial topology.
It is found that a higher dimensional universe with a product space
with primordial black holes pair is less probable to nucleate when the
extra dimensions scale factors do not vary in an inflating universe. \\

PACS number(s) : 04.20.Jb, 04.60.+n, 98.80.Hw
\end{abstract}

\pagebreak

\section{INTRODUCTION : }

Considerable work has been witnessed in the last two decades on
the problem of the creation and subsequent evolution of primordial black 
holes in different contexts. In quantum cosmology one of the most challenging
problems is the existence of primordial black holes ( hence forth, PBH ). It
is now understood that the
black holes are formed either, (i) due to gravitational collapse of a massive
 star when the  mass of a  given star exceeds about twice that of 
the solar mass or (ii) due to quantum fluctuation of matter distribution in 
the early universe.
In 1975, the remarkable discovery of Hawking radiation ushered in a new era in
black hole physics [1]. The life of the first kind of black holes are long, 
which are comparable
to the age of the universe. Therefore, the only hope of confirming Hawking 
radiation by
observation is through PBH hunting.  Recently, Bousso and Hawking [2] 
( in short, BH ) calculated the probability of the quantum creation of 
a universe 
with a pair of primordial black holes in  (3 + 1) dimensional universe. 
In the paper they considered two different 
euclidean space-time : (1) a universe with space-like sections with 
topology $S^{3}$ and (2) a universe with space-like section with topology 
$ S^{1} \times S^{2}$ , as is obtained in the Schwarzschild- de Sitter solution.
With a suitable choice of the time variables, the first one  gives an 
inflationary (de Sitter) universe while the second describes a Nariai 
universe [3], an inflationary universe with a pair of black holes.  
BH considered in their paper a massive scalar field which provided an 
effective cosmological constant for a while through a slow-rolling potential 
(mass-term).
Paul {\it et al.} [4] following the approach of BH studied the probability
of creation of PBH including $R^{2}$-term in the Einstein action and found
that the probability is very much suppressed in the $R^{2}$ -theory. However,
the evolution of a universe at  Planck time t $\sim$ $M_{P}^{-1}$ ,  may be
better understood in
 the framework of quantum gravity. A consistent quantum theory of gravity
 is yet to emerge. Kaluza and Klein [5] first initiated the study to 
formulate a unified theory of gravity by introducing an extra dimension. The 
approach has been revived and considerably generalized after realising that 
many interesting theories of 
particle interactions need more than four dimensions for their 
formulation [6].
It is considered essential to check if consistent cosmological or 
astrophysical solution, which can accommodate, these theories are also 
allowed. In this paper we intend to calculate the probability of 
quantum creation 
of a higher dimensional universe with $ R \times S^{D-1} $ topology and 
$R \times S^{1} \times S^{D-2}$ topology  which accommodates
a pair of primordial black holes. 
To calculate the probabilities for the creation of a universe with the
two types of topology
we use a semiclassical approximation for the evaluation of  the 
euclidean path integrals. The condition that a classical spacetime 
should emerge, to a good 
approximation, at a large Lorentzian time  was ensured by a choice of the
 path of the time parameter $\tau$ along the $\tau^{Re}$ axis 
from 0 to $\frac{\pi}{2H}$ and 
 then continues along the $\tau^{im}$ axis. 
The semiclassical approximation for the wave-function of the universe is given by
\begin{equation}
\Psi_{o} [ h_{ij}  , \Phi_{\partial  M} ]   \approx \sum_{n} A_{n} e^{- I_{n}}
\end{equation}
where the sum is over the saddle points of the path integral, 
and $I_{n}$
denotes the corresponding Euclidean action.  The probability measure of the
creation of PBH is
\[
P[ h_{ab} , \Phi_{\partial M} ] \sim  e^{ - 2 \;  I^{Re}}
\]
where $h_{ab}$ is the boundary metric and  $I^{Re}$  is the real part of
the action corresponding to the dominant saddle point, i.e. the classical
solution satisfying the Hartle-Hawking [ henceforth, HH ] boundary
conditions [7]. It is also important to explore the  possible causes of 
splitting of a multidimensional universe. It has been shown that
the observed  3 physical space may
 obtained starting from a multidimensional universe by the Freund-Rubin ansatz
 [8] . Also Candelas and Weinberg [9] observed  that  a split universe 
may be  obtained by considering
 fluctuations
in massless ( N + 1)  dimensional matter fields. In this paper we show that 
the existence
of a pair of PBH  requires a split universe. However, the probability of 
creation
of such a universe is found to be very low. It is found that a higher dimensional
universe nucleates with all of its dimensions expanding equally. However, in course
of its evolution, the universe with 3-physical space behave differently from 
that of the internal space [10]. It was believed that all inflationary models lead
to $ \Omega_{o} \sim 1$ to a great accuracy. This view was modified after it
was discovered  that there is a special class of inflaton effective potentials which
may lead to a nearly homogeneous open universe with  $ \Omega_{o} \leq 1$ 
at the present epoch. Cornish {\it et al.} [11, 12] studied the problem of pre-inflationary
homogeneity and outlines the possibility of creation of a small, compact, negatively
curved universe. We show that a universe with $S^{D}$-topology may give 
birth to
an open inflation.

The paper is organised as follows : in sec. 2 we  write the gravitational
 action for
a multidimensional universe and present gravitational instanton solutions and
in sec. 3 we use the action to estimate the relative probability of the 
two types of the universes and in sec. 4 we give a brief discussion.

\section{ GRAVITATIONAL INSTANTON SOLUTIONS WITH OR WITHOUT 
A PAIR OF PRIMORDIAL BLACK HOLES: }

We consider a higher dimensional Euclidean action given by
\[
I_{E} =  - \frac{1}{16\pi} \int d^{D} x \sqrt{g} [ R  - 2 \Lambda ]
 - \frac{1}{8\pi} \int_{\partial M} d^{D-1} x \sqrt{h} K 
\]
where g is the D-dimensional Euclidean metric, R is the Ricci scalar
 and $\Lambda$ is the D-dimensional cosmological constant.
In the gravitational surface term, $h_{ab}$ is the boundary metric and 
$K = h^{ab} K_{ab}$ is the trace of the second fundamental form of the boundary
$\partial M$ in the metric. The Euclidean action takes the form
\begin{equation}
I_{E} =  - \frac{1}{16\pi} \int d^{D - 1} x \sqrt{h} [ K_{ab} K^{ab} + K^{2}
 + ^{D - 1}R  - 2 \Lambda ]
 + \frac{1}{8\pi} \int_{\tau = 0} d^{D-1} x \sqrt{h} K 
\end{equation}
using the no boundary proposal of Hartle-Hawking which is assumed to satisfy
at $\tau = 0$. Here $^{D - 1}R$ is the scalar curvature of the (D - 1) 
dimensional surface.

{\large (A)  Topology $S^{D}$, the de Sitter spacetime :}

In this section we study vacuum solutions of the Euclidean Einstein equation 
with a cosmological constant in D dimensions. We now look for a solution 
with spacelike section $S^{n}$. Hence we choose the
D-dimensional metric ansatz
\begin{equation}
ds^{2} = d\tau^{2} + a^{2} (\tau) d\Omega_{n}^{2}
\end{equation}
where $a$ is the scale factor of a D-dimensional universe,  $ n = D - 1 $
and $ d\Omega_{n}^{2} $ is a line element on the unit (D - 1) sphere.
The scalar curvature is given by
\[
R = -  \left[ 2 n \frac{\ddot{a}}{a} + n (n - 1) \left(
    \frac{\dot{a}^{2}}{a^{2}} - \frac{1}{a^{2}} \right) \right] .
\]
where an overdot denotes differentiation with respect to $\tau$. We rewrite 
the action (2) and obtain 
\begin{equation}
I_{E} = -  V_{o} \int_{\tau = 0}^{\tau_{\partial M}} 
         \left[ 2 (D - 1)(D - 2)
          a^{D - 3} ( \dot{a}^{2} + 1 ) - 2 \Lambda a^{D - 1} \right] d \tau 
      + [ - 2 V_{o} (D - 1) \dot{a}
       a^{D - 2}]_{\tau = 0} 
\end{equation}
where $V_{o} = \frac{1}{16 \pi} 
 \frac{2 \pi^{(D + 1)/2}}{\Gamma ( \frac{D + 1}{2})}$, 
    we have eliminated $\ddot{a}$ by integration by parts. The field 
equations can now be obtained by varying $I_{E}$ with respect to $a$ , 
giving
\begin{equation}
2 \frac{ \ddot{a}}{a} + (D -3) \left[ 
     \frac{ \dot{a}^{2} - 1}{a^{2}} \right] 
   +  \frac{2 \Lambda}{D - 2} = 0 .
\end{equation}
The field eq.(5) allows an instanton solution which is given by
\begin{equation}
a = \frac{1}{H_{o}} \; sin  H_{o} \tau 
\end{equation}
where
$ H_{o}^{2} = \frac{ 2 \Lambda }{(D - 1)(D - 2)} $.
We note that this solution satisfies the HH no boundary conditions viz., 
$a(0) = 0 $, $ \dot{a} (0) = 1$. One can choose a path along the $\tau^{Re}$
axis to  $\tau = \frac{\pi}{2 H}$, the solution describes half of the Euclidean
de Sitter instanton $S^{D}$.  Analytic continuation of the metric (3) to 
Lorentzian region, $x_{1} \rightarrow \frac{\pi}{2} + i \sigma $, gives
\begin{equation}
ds^{2} = d \tau^{2} + a^{2}(\tau) [ - d\sigma^{2} + \cosh^{2} \sigma
 d \Omega_{n-1}^{2} ]
\end{equation}
which is a spatially inhomogeneous de Sitter like metric. However, if one sets $
\tau = i t $ and  $ \sigma = i \frac{\pi}{2} + \chi $, the metric becomes
\begin{equation}
ds^{2} = -  dt^{2} + b^{2}(t) [ d\chi^{2} + \sinh^{2} \chi
 d \Omega_{n-1}^{2} ]
\end{equation}
where $b(t) = - i a( it )$. It describes the creation of an open inflationary higher
dimensional universe.
The real part of the Euclidean action 
corresponding to the solution calculated by following the complex contour
of $\tau $ suggested by BH is given by
\begin{equation}
I^{Re}_{S^{D - 1}} = - 2  V_{o} 
\sqrt{ \frac{ ( (D - 1)(D - 2) )^{D}}{ ( 2 \Lambda )^{D - 2}}}
       \left[ I_{n - 2} - I_{n} \right]
\end{equation}
where $n = D - 1$  and
 $ I_{n} = \int^{ \frac{\pi}{2}}_{0} \; sin^{n} y \; dy  $  with
$y = H_{o} \tau $. The value of the integral $I_{n}$ 
depends on the dimensions of the universe, for odd number of dimensions
\[
I_{n} = \frac{n - 1}{n}  \frac{n - 3}{n - 2} .... 
\frac{3}{4} \frac{1}{2} \frac{\pi}{2} , \; \;  for \; n \; even
\]
and for even number of dimensions 
\[
\; \; \; = \frac{n - 1}{n} \frac{n - 3}{n - 2} ....
 \frac{4}{5}  \frac{2}{3} 1 ,  \; \; for \; n \; odd.
\]
With the chosen path for $\tau $ ,  the solution describes 
half the de Sitter Instanton in a higher dimensional universe with $S^{D}$ 
topology, joined to a real Lorentzian hyperboloid of
topology  $R^{1} \times S^{D - 1}$. It can be joined to any boundary satisfying 
the condition 
$a_{\partial M} > 0$ . 
For $a_{\partial M} > H_{o}^{- 1}$ , the wave function oscillates and 
predicts a classical
space-time. 

{\large (B) Topology $ S^{1} \times S^{d} $ :}

We consider vacuum solution of the Euclidean Einstein equation
with a cosmological constant and look for a universe with $S^{1} \times
S^{d}$ -spacelike sections as this topology accommodates a pair of black 
holes. The corresponding ansatz for (1 + 1 + d) dimensions is given by
\begin{equation}
ds^{2} = d \tau^{2} + a^{2}(\tau) dx^{2} + b^{2}(\tau) \;  d \Omega_{d}^{2} 
\end{equation}
where $a( \tau ) $ is the scale factor of two sphere and $b( \tau )$ is the
scale factor of the d-sphere given by the metric 
\[
d\Omega_{d}^{2} = dx_{1}^{2} + sin^{2} x_{1} \; dx_{2}^{2} +
 sin^{2} x_{1} \; sin^{2} x_{2} \;  dx_{3}^{2} +  \;  ...  \; \;  d-space .
\]
The scalar curvature is given by
\[
R = - \left[ 2 \frac{\ddot{a}}{a} +  2d \frac{\ddot{b }}{b}
     + d(d - 1) \left( \frac{\dot{b}^{2}}{b^{2}} - 
   \frac{1}{b^{2}} \right) + 2d \frac{\dot{a} \dot{b}}{a b} \right] .
\]
The Euclidean action (2) becomes
\[
I_{E} = - V_{o}' \int_{\tau = 0}^{\tau_{\partial M}} 
 \left[  d(d - 1)a b^{n - 2} \left( \dot{b }^{2} + 1 \right)
         + 2 d \dot{a} \dot{b} b^{d - 1} - 2 \Lambda a b^{d}
     \right] d \tau  
 + V_{o}' [ - 2 \dot{a} b^{d} - 2 d a 
\dot{b} b^{d - 1}]_{\tau = 0}
\]
where $ V_{o}' = \frac{1}{4 \pi}
 \frac{ \pi^{(D - 1)/2}}{\Gamma ( \frac{D - 1}{2})}$.

Variation of the action with respect to the scale factors $a $ and $b$ gives
the following field equations
\begin{equation}
 2d \frac{\ddot{b}}{b} + d(d - 1) \left(  \frac{\dot{b}^{2}}{b^{2}} - 
   \frac{1}{b^{2}} \right) + 2 \Lambda  = 0
\end{equation}
\begin{equation}
\frac{\ddot{a}}{a} +  (d - 1) \frac{\ddot{b}}{b} + \frac{(d - 1)(d - 2)}{2}
 \left( \frac{\dot{b}^{2}}{b^{2}} - 
   \frac{1}{b^{2}} \right) + (d -1) 
\frac{\dot{a} \dot{b}}{ a b} + \Lambda  = 0 .
\end{equation}
The field eqs. (11) and (12) admit a solution which is given by
\[
a = \frac{1}{H_{o}} \; sin ( H_{o} \tau ) , \; \; \;
b = \sqrt{d - 1} \; H_{o}^{- 1} ,
\]
\begin{equation}
H_{o}^{2} = \frac{2}{d} \; \Lambda
\end{equation}
for $n > 1$ i.e., valid for dimensions $D > 2 + 1$.
This solution satisfies the HH boundary conditions
$a(0) = 0 $, $ \dot{a} (0) = 1,  b(0) = b_{o} $, $ \dot{b} (0) = 0$.
Analytic continuation of the metric ( 10 ) to Lorentzian region, i.e.,
$\tau \rightarrow it $ and $ x \rightarrow \frac{\pi}{2} + i \sigma $ yields
\begin{equation}
ds^{2} = -  dt^{2} + c^{2}(t) d\sigma^{2} + H_{o}^{-2} d \Omega_{n-1}^{2}
\end{equation}
where $c(t) = - i a( it )$. In this case the analytic continuation of time and space
do not give an open inflationary universe. The corresponding Lorentzian solution
is given by
\[
a(\tau^{Im}) |_{\tau^{Re} = \frac{\pi}{2H}} = H^{-1} \cosh H \tau^{Im} ,
\]
\[
b(\tau^{Im}) |_{\tau^{Re} = \frac{\pi}{2H}} = H^{-1} 
\]
Its space like sections can be visualised as  (D - 1) spheres of radius $a$
with a `hole' of radius $b$ punched through the north and south poles. The
physical interpretation of the solution is that of (D - 1) - spheres containing
two black holes at opposite ends. The black holes have the radius $H^{-1}$
accelerates away from each other with the expansion of the universe.
The
real part of the action can now be determined following the contour
 suggested  by BH [2], and it is given by
\begin{equation}
I^{Re}_{S^{1} X S^{d}} = - \left[ 2 V_{0}' \left( \frac{d(d -1)}{2 \Lambda}
\right)^{d/2} \right]
\end{equation}
where $d = D - 2$.
The solution (13) describes a universe with two black holes at the poles
of a (D - 1)- sphere. The contribution of the action in this case comes from
the surface term only.

\section{EVALUATION FOR THE PROBABILITY FOR PRIMORDIAL BLACK HOLES :}

In the previous section we have calculated the actions for inflationary 
universe with or without a pair of black holes. We now compare the probability
measure.
The probability for creation of a higher dimensional de Sitter universe 
may be obtained from the action (9).  Thus, the probability for
nucleation of a higher dimensional universe
without PBH is given by
\begin{equation}
P_{S^{D - 1}} \sim \; Exp \left[ \frac{(D - 1)(D - 2)}{4 \pi} .
       \frac{2 \pi^{ (D + 1)/2}}{ \Gamma (\frac{D + 1}{2})} 
    \left( \frac{(D - 1) (D - 2)}{2 \Lambda} \right)^{ \frac{D - 2}{2}}
      \left(I_{D - 3} - I_{D - 1} \right) \right] 
\end{equation}
where $I_{n} < 1$ and $ I_{D - 3} > I_{D - 1}$ . The action evaluated in this
case is always
 negative.
However for an inflationary universe with a pair of black holes the 
corresponding probability of nucleation can be obtained from the
action (15).  The corresponding probability is
\begin{equation}
P_{S^{1} \times S^{D-2}} \sim Exp \left[  
   \frac{ \pi^{(D - 1)/2}}{  2 \Gamma (\frac{D - 1}{2})} 
\left( \frac{(D - 2)(D -3)}{2
   \Lambda} \right)^{(D - 2)/2} 
      \right]
\end{equation}
for $D > 3$ . For $D = 4$ we recover the result obtained by 
Bousso and Hawking
[2]. However, in a higher dimensional universe the probability of
 nucleation of a $S^{D -1}$ - topology is found more 
than that of a universe with
 $S^{1} \times S^{D - 2}$ - topology in the presence of 
primordial black holes. 

\section{ DISCUSSIONS : }

In this work we have evaluated the probability for primordial
black holes pair creation in a higher dimensional universe ( $D > 4$ ). In
section 2, we have obtained the gravitational instanton solutions in two 
cases : a universe with (i) $R \times S^{D - 1}$ - topology  and a universe with
(ii) $ R \times S^{1} \times S^{D - 2} $ -topology respectively. The 
Euclidean  action is then evaluated. We found  that the probability of a universe
with $R \times S^{D - 1} $ topology turns out to be much lower than a universe with
a split universe with topology $ R \times S^{1} \times S^{D - 2} $ to begin with.
It may be mentioned here that one gets a regular instanton in $S^{D} $ - topology
if there are no black holes. The existence of black holes  
restricts such a regular topology. 
The result obtained here on the probability of creation of a higher dimensional
universe with a  pair of primordial black holes is found to be strongly 
suppressed. Such a universe  inflates in one particular dimensions whereas 
the other scale factors do not vary with time. It may be mentioned here that
in a de Sitter like 
multidimensional universe all the scale factors expands equally to begin with.
However, analytic continuation of a $R \times S^{D - 1}$ metric considered here
to Lorentzian region leads to an open 3 - space and closed extra space. One
obtains Hawking-Turok [13,14] type open inflationary universe in this case. In the 
other type of topology with split universe such a open inflation is not permitted.
A detail study of an open inflationary universe will be discussed elsewhere.
Finally the result obtained here may be used to answer the question on 
splitting
of a multi dimensional
universe. 

{\large {\it Acknowledgement  :}}

I like to thank 
the Inter-University University Centre for Astronomy
and Astrophysics ( IUCAA ), Pune for warm hospitatlity  during a visit for 
the work. I also thank the referee for his valuable suggestions 
and constructive criticism that has resulted in an improvment
in presentation of the work.

\pagebreak

\end{document}